\documentclass[a4paper,11pt]{article}
\usepackage{pos}

\usepackage{float}
\restylefloat{table}
\usepackage{hyperref}
 \usepackage{bm}

\usepackage{orcidlink}
\usepackage{diagbox}
\usepackage{makecell, multirow}
\usepackage{pgf}
\usepackage{soul}
\usepackage{graphicx, color}
\usepackage{rotating}

\allowdisplaybreaks

\definecolor{link_blue}{RGB}{51,102,204}

\newcommand{\etal}{\textit{et al.}}

\hypersetup{%
pdftitle = {title},
pdfsubject = {},
pdfkeywords = {},
colorlinks = {true},
filecolor = {black},
linkcolor = {link_blue},
menucolor = {black},
citecolor = {link_blue},
urlcolor = {link_blue},
}{}

\title{Heavy baryons with relativistic quarks}
\ShortTitle{Heavy baryons with relativistic quarks}

\author*[a,b]{Archana Radhakrishnan~\orcidlink{0000-0001-9357-1360}}
\author[a]{Debsubhra Chakraborty~\orcidlink{0000-0001-5815-4182}}
\author[a]{Nilmani Mathur~\orcidlink{0000-0003-2422-7317}}

\affiliation[a]{Department of Theoretical Physics, Tata Institute of Fundamental Research,\\Homi Bhabha Road, Mumbai 400005, India}
\affiliation[b]{Centre for High Energy Physics, Indian Institute of Science,\\Bangalore 560012, India}

\emailAdd{archana@theory.tifr.res.in}
\abstract{
We present a lattice QCD study of heavy baryons containing charm and bottom quarks, with particular emphasis on the relativistic treatment of all valence quarks.
We use $N_f=2+1+1$ HISQ ensembles at the physical point to compute ground-state
energies of spin-$3/2^+$ baryons, including singly-, doubly-, and triply-heavy charmed and bottom baryons.
This work represents the first investigation of heavy baryons using {\it fully relativistic bottom quarks}.}

\FullConference{The 42nd International Symposium on Lattice Field Theory (LATTICE2025)\\
2-8 November 2025\\
Tata Institute of Fundamental Research, Mumbai, India\\}

\renewcommand{\hookAfterAbstract}{%
\par\bigskip
\textsc{Preprint}: TIFR/TH/26-14
}

\begin{document}
\maketitle

\section{\label{sec:intro}Introduction}
Heavy baryons provide a clean theoretical environment for studying nonperturbative QCD dynamics, particularly in systems where light-quark degrees of freedom play a subdominant role. In recent years, experimental progress, most notably the discovery of the $\Xi_{cc}^{++}$ by LHCb~\cite{Aaij:2017ueg}, has renewed interest in the study of hadrons with heavy quarks. With the increasing luminosities in experiments, particularly at LHCb, BESIII and Belle II, the lattice calculation of spectra of baryons with bottom and charm quarks is timely.

Lattice QCD is currently playing a very important role in our theoretical understanding of the QCD spectra.
An example of this is seen in the prediction of doubly charmed baryon, $\Xi_{cc}^{++}$, before its discovery. Most lattice calculations consistently predict a mass near
3620~MeV~\cite{PhysRevD.64.094509, PhysRevD.66.014502, Brown:2014ena}, in agreement with the LHCb observation of $\Xi_{cc}^{++}$, the isospin partner of the SELEX candidate. The SELEX experiment claimed the existence of the doubly-charmed baryon ($\Xi_{cc}^+$)~\cite{Mattson:2002vu} at 3519~MeV but was not established by subsequent experiments.

In this work, we focus on the spectroscopy of spin-$3/2^+$ heavy baryons, including
systems with one, two, and three charm and/or bottom quarks. Such states are experimentally
challenging to access currently, yet
they provide a theoretically clean probe of heavy-quark dynamics in QCD. With experiments striving for higher luminosities such baryons could even be accessible to experiments in future.

The key novelty of this study is the use of a fully relativistic formulation for all valence quarks, including bottom. This allows charm and bottom quarks to be treated
on equal footing and avoids the systematic limitations inherent to effective field theory approaches such as NRQCD. This is the {\it first} lattice study of heavy baryons employing relativistic bottom quarks.

\section{\label{sec:setup}Lattice Setup}

Our calculations are performed on MILC $N_f=2+1+1$ gauge ensembles generated with the
HISQ action~\cite{Bazavov:2012xda}. The valence strange, charm, and bottom quarks are
also implemented using the HISQ action. Exact isospin symmetry is assumed.

\begin{figure}[h!]
    \centering
    \includegraphics[scale=0.29]{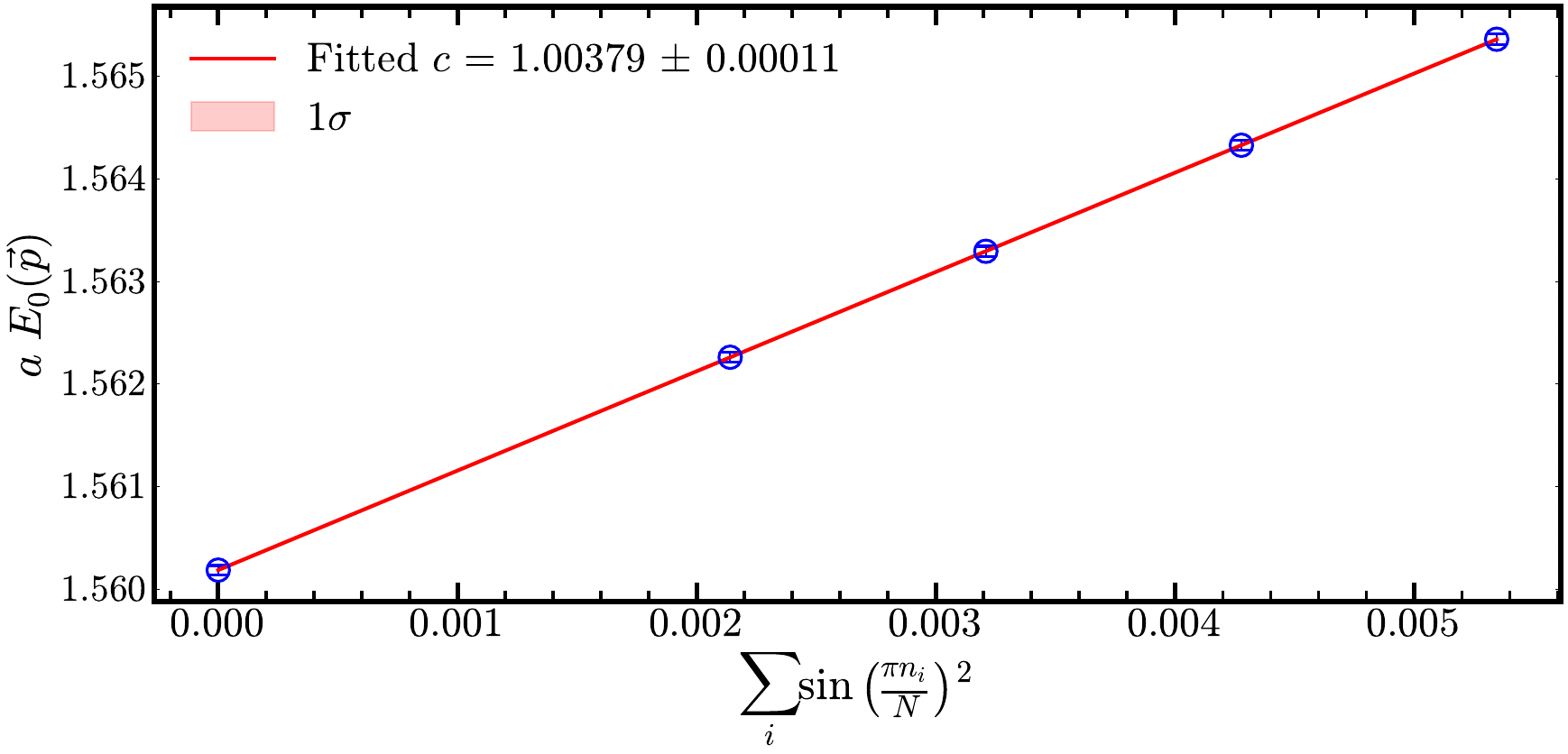} ~~
    \includegraphics[scale=0.29]{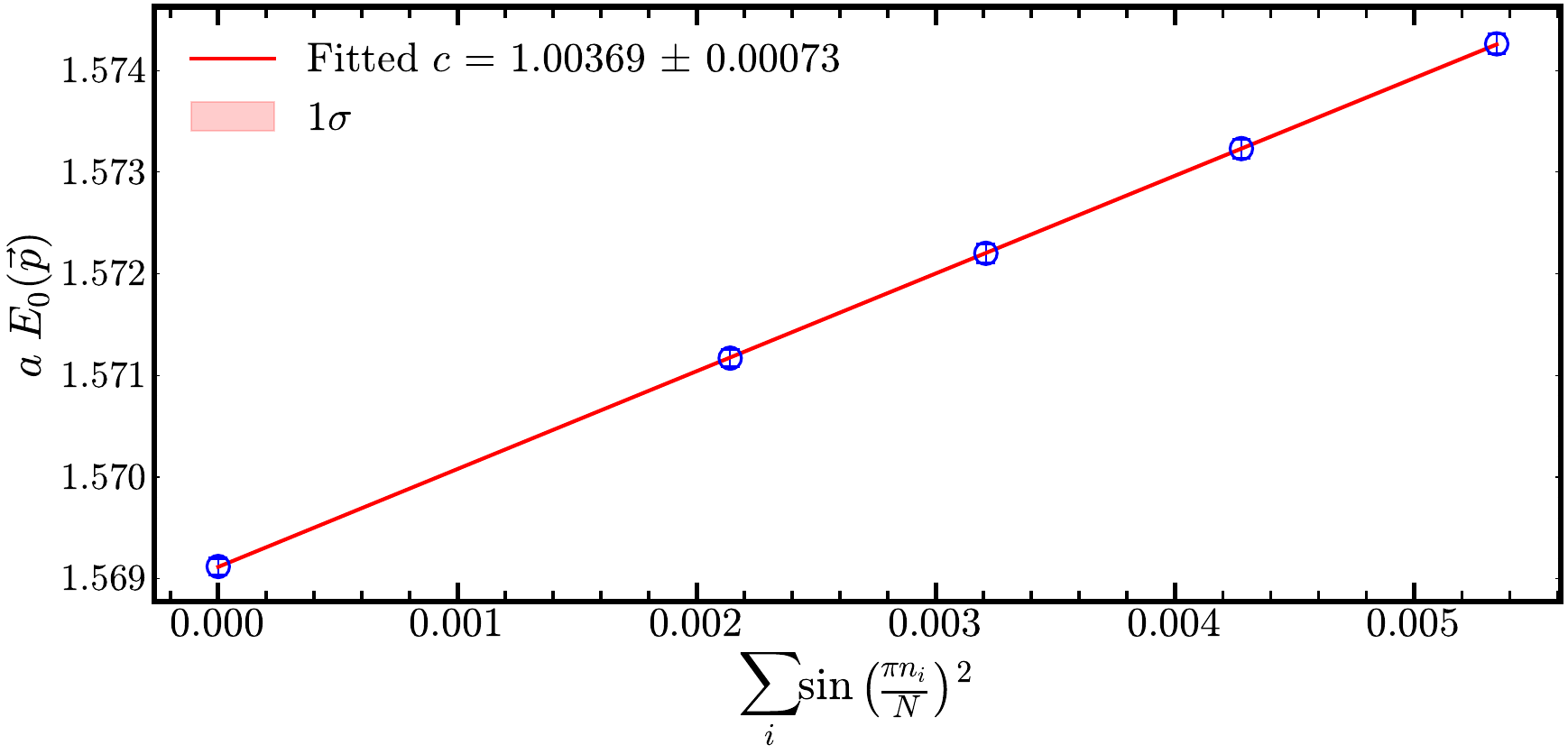}
    \caption{Dispersion relation of $\eta_b$ (top) and $\Upsilon$ (bottom). The ground state energy $aE_0(\vec{p})$ in lattice units is plotted against the lattice momentum squared parameter $\sum_i \sin\left(\frac{\pi n_i}{N}\right)^2$. The red line represents a fit to the data with a $1\sigma$ error band, yielding slope parameters $c = 1.00379 \pm 0.00011$ ($\eta_b$) 
and $c = 1.00369 \pm 0.00073$ ($\Upsilon$). A fitted value of $c \approx 1$ demonstrates the successful tuning of 
the bare bottom quark mass, confirming that the correct relativistic 
behavior is restored on the lattice.}
    \label{fig:disp}
\end{figure}

The ensemble used in this analysis has lattice spacing,
$a = 0.0327~\mathrm{fm}$,
with spatial volume $L^3 = 96^3$ and temporal extent $T=288$. The fine lattice spacing
allows for controlled discretization errors even for heavy bottom quarks.
\begin{table}[ht]
\centering
\begin{tabular}{cccccc}
\hline
$a$ (fm) & $L^3\times T$ & $am_l$ & $am_s$ & $am_c$ & $am_b$ \\
\hline
0.0327 & $96^3\times288$ & 0.00223 & 0.01115 & 0.1316 & 0.6223 \\
\hline
\end{tabular}
\caption{Parameters of the $N_f=2+1+1$ HISQ ensemble of MILC used in this work.}
\label{tab:ensemble}
\end{table}
Using the lattice spacings from~\cite{Hatton:2020qhk}, the charm mass was previously tuned in~\cite{Dhindsa:2024erk}. 
The bottom mass is tuned by equating the spin-averaged kinetic
mass of the $\overline{1S}$ bottomonium states to the physical value.  The validity of the tuning is verified by extracting the dispersion relations of the $\eta_b$ and $\Upsilon$ mesons, which exhibit the correct momentum dependence Fig.~\ref{fig:disp}.

\section{Relativistic Treatment of Bottom Quarks}

Traditional lattice computations of bottom-hadron spectra employ Non-Relativistic QCD (NRQCD), an effective field theory constructed via an expansion in the heavy quark velocity squared, $v^2$. The heavy quark dynamics are governed by an effective Hamiltonian of the form:
\begin{equation}
    H_{\text{eff}} = H_0 + \delta H, \quad H_0 = -\frac{\bm{\Delta}^{(2)}}{2M_0},
\end{equation}
where $\delta H$ includes higher-order spin-dependent and relativistic corrections. However, as a non-renormalizable effective theory, NRQCD suffers from severe power-law divergences in radiative corrections. The matching coefficients $C_k$ for higher-dimension operators exhibit behavior proportional to inverse powers of the lattice mass:
\begin{equation}
    C_k(aM_0) \sim \sum_{n,l} \alpha_s^l \, (aM_0)^{-n}.
\end{equation}
where $l$ denotes the order in perturbation theory. These divergences preclude a true continuum limit ($a \to 0$), as the coefficients blow up when the lattice cutoff exceeds the heavy mass scale. Consequently, simulations are restricted to coarse lattices where $aM_0 \gtrsim 1$, introducing irreducible systematic uncertainties from the truncation of the velocity expansion.

In contrast, the use of a relativistic action can avoid such problems. However, employing a relativistic quark action for bottom baryons requires a very small lattice spacing to suppress discretization errors 
$am \ll 1$. Thanks to the superfine lattices generated by the MILC Collaboration, in this work we employ the Highly Improved Staggered Quark (HISQ) action  \cite{Follana:2006rc} for the bottom quark sector, which satisfies these requirements. This fully relativistic formalism preserves a remnant $U(1)_\epsilon$ chiral symmetry, protecting the theory from odd-power discretization errors ($\mathcal{O}(a), \mathcal{O}(a^3)$). 
Furthermore, for heavy quarks, the HISQ action along with the mass-dependent tuning of the coefficient of the Naik term~\cite{Follana:2006rc} eliminates not only the tree-level $\mathcal{O}(a^2)$ errors in the Dirac derivative, but also the leading tree-level $\mathcal{O}((am_h)^4)$ errors, which is a significant correction given our bottom quark mass of $am_b = 0.6223$.

Unlike NRQCD, the HISQ action is renormalizable and permits a rigorous continuum extrapolation of observables for heavy hadrons. By generating data at multiple lattice spacings (including fine lattices where $am_b < 1$), discretization artifacts can be systematically removed using the functional form expected from Symanzik effective theory. 
Furthermore, utilizing the same lattice action for light ($u,d,s$), charm ($c$), and bottom ($b$) quarks ensures a unified treatment, avoiding any mixed action effects \cite{PhysRevD.86.014501, Basak:2012py}.
This uniformity facilitates the exact cancellation of common lattice artifacts in mass ratios, leading to significantly reduced systematic errors in heavy-baryon spectroscopy.

\section{\label{sec:meth}Methodology}

Hadron masses are extracted from the asymptotic time dependence of Euclidean two-point correlation functions. In the staggered fermion formalism, temporal correlators for baryons receive contributions not only from the desired physical states but also from their opposite-parity partners, which appear as $(-1)^t$  oscillating contributions in Euclidean time. Because our analysis relies on a single interpolating operator per channel, the spectral decomposition of the two-point correlator takes the exact analytic form:
\begin{equation}
    C(t) = \sum_{n=0}^{N-1} A_n e^{-M_n t} + \sum_{n=0}^{\tilde{N}-1} \tilde{A}_n (-1)^t e^{-\tilde{M}_n t},
    \label{eq:correlator}
\end{equation}
where $M_n$ and $A_n$ represent the masses and spectral overlaps of the normal-parity states, while $\tilde{M}_n$ and $\tilde{A}_n$ denote the masses and amplitudes of the oscillating opposite-parity states. 

Lacking a broad operator basis, in this work we are unable to employ a matrix diagonalization approach (such as the Generalized Eigenvalue Problem) to project out excited states. Instead, the ground-state mass $M_0$ for a baryon is isolated via direct, high-precision multi-exponential fits to Eq.~\eqref{eq:correlator}. To ensure the stability of these fits and suppress excited-state contamination, temporal smoothing techniques are applied to the raw correlators~\cite{DeTar:2014gla}. This systematically dampens the high-frequency $(-1)^t$ oscillatory components prior to numerical minimization. The ground-state parameters are fixed when the correlated $\chi^2 / \text{d.o.f.}$ is minimized and the extracted mass is stabilized with a clear plateau over a carefully chosen temporal window $[t_{\text{min}}, t_{\text{max}}]$.

We also examine the residual taste-symmetry breaking inherent to staggered fermions. The HISQ action substantially suppresses one-loop $\mathcal{O}(\alpha_s a^2)$ taste-exchange interactions~\cite{Follana:2006rc}. Consistent with recent high-precision HISQ analyses of $\Omega_{ccc}$~\cite{Dhindsa:2024erk}, finite-lattice-spacing taste splittings are found to be heavily suppressed and remain consistent with zero within our current statistical uncertainties.

\section{\label{sec:res}Results}
We extract ground-state energies for spin-$3/2^+$ baryons with flavor content
$c c c$, $s s c$, $c c s$, $s s b$, $s c b$, $c c b$, $s b b$, $c b b$ and $b b b$. 

Our primary results for the ground-state masses of the heavy $\Omega$ baryons are summarized and compared with previous lattice QCD determinations and experimental data. By employing a highly improved, fully relativistic HISQ framework for all quark flavors, we achieve a determination of the heavy baryon spectrum across both the charm and bottom sectors. 

\subsection{Strange-Charmed Baryon Spectra}
\begin{figure}[htbp]
    \centering
    \includegraphics[width=0.9\textwidth]{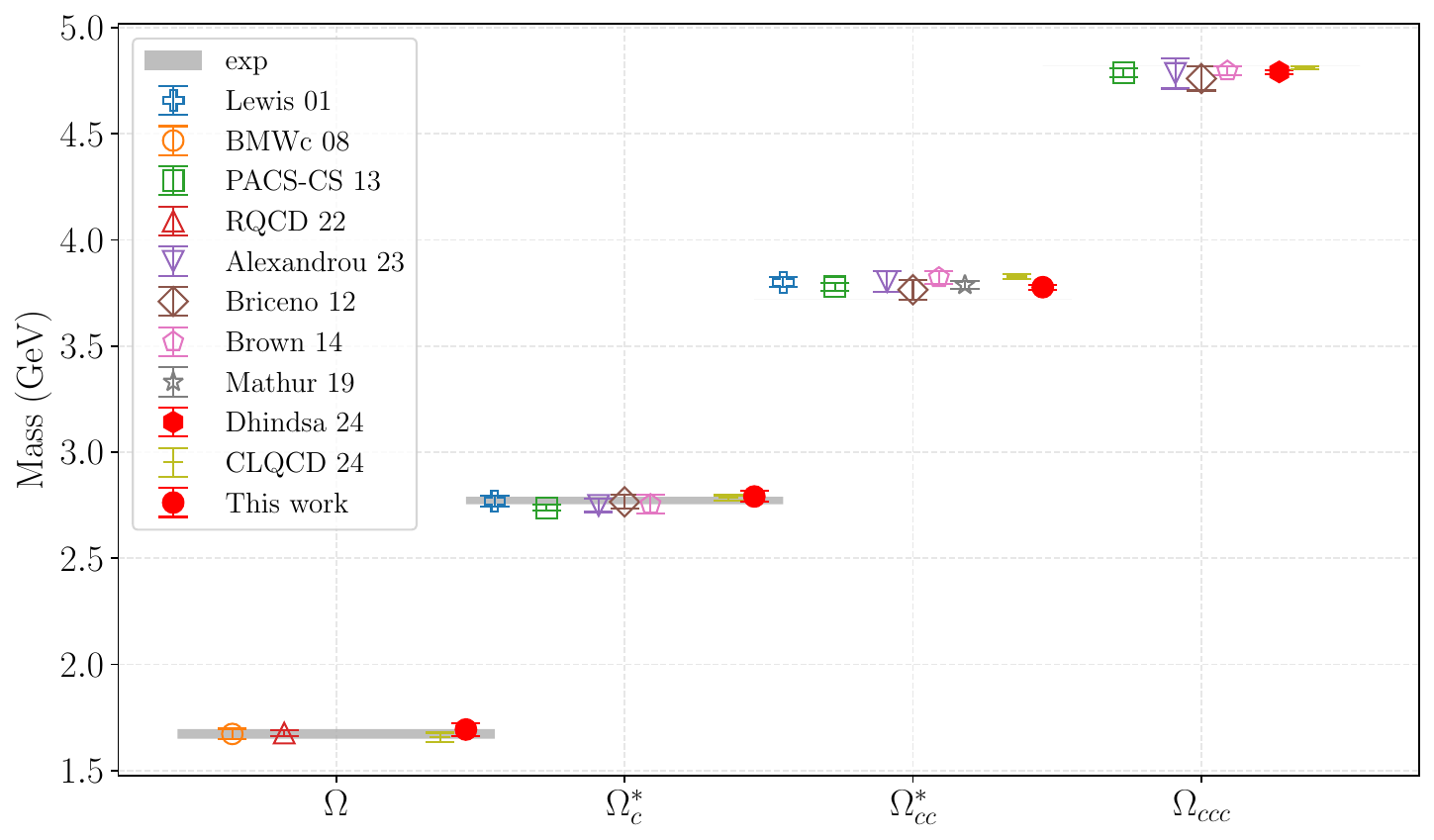}
    \caption{\label{fig:charm_summary} 
    Summary of the ground-state masses for the charmed $\Omega$ baryon family ($\Omega$, $\Omega_c^*$, $\Omega_{cc}^*$, $\Omega_{ccc}$). Our fully relativistic HISQ results are shown as red circles. Experimental values from the PDG are denoted by gray horizontal bands. Previous lattice QCD determinations utilizing various fermion actions are included for comparison.}
\end{figure}

Figure~\ref{fig:charm_summary} presents our extracted ground state masses for the strange-charm $\Omega$ baryon family ($\Omega$, $\Omega_c^*$, and $\Omega_{cc}^*$), plotted alongside experimental values, where available, and selected lattice results from the literature. For the states with established experimental measurements ($\Omega$ and $\Omega_c^*$), our results, indicated by the open red circles, postdict the physical values (horizontal gray bands) with good accuracy. 

For the doubly charmed state ($\Omega_{cc}^*$), which currently lacks definitive experimental observation, our results serve as a predictive benchmark. Our determination of the $\Omega^*_{cc}$ mass is in agreement with the previous lattice calculations ~\cite{
PhysRevD.64.094509, PhysRevD.66.014502, Brown:2014ena,
Briceno:2012wt, Mathur:2018rwu,Alexandrou:2023dlu,RQCD:2022xux,Hu:2024mas,BMW:2008jgk}. 
The $\Omega_{ccc}$ mass shown in 
Fig.~\ref{fig:charm_summary} is taken from our earlier dedicated 
study~\cite{Dhindsa:2024erk}, using a similar setup, where agreement with previous lattice 
determinations was demonstrated. Our prediction for the  mass of $\Omega_{ccc}$ baryon was most precise to date.

\subsection{Bottom Baryon Spectra}
\begin{figure}[htbp]
    \centering
    \includegraphics[width=0.7\textwidth]{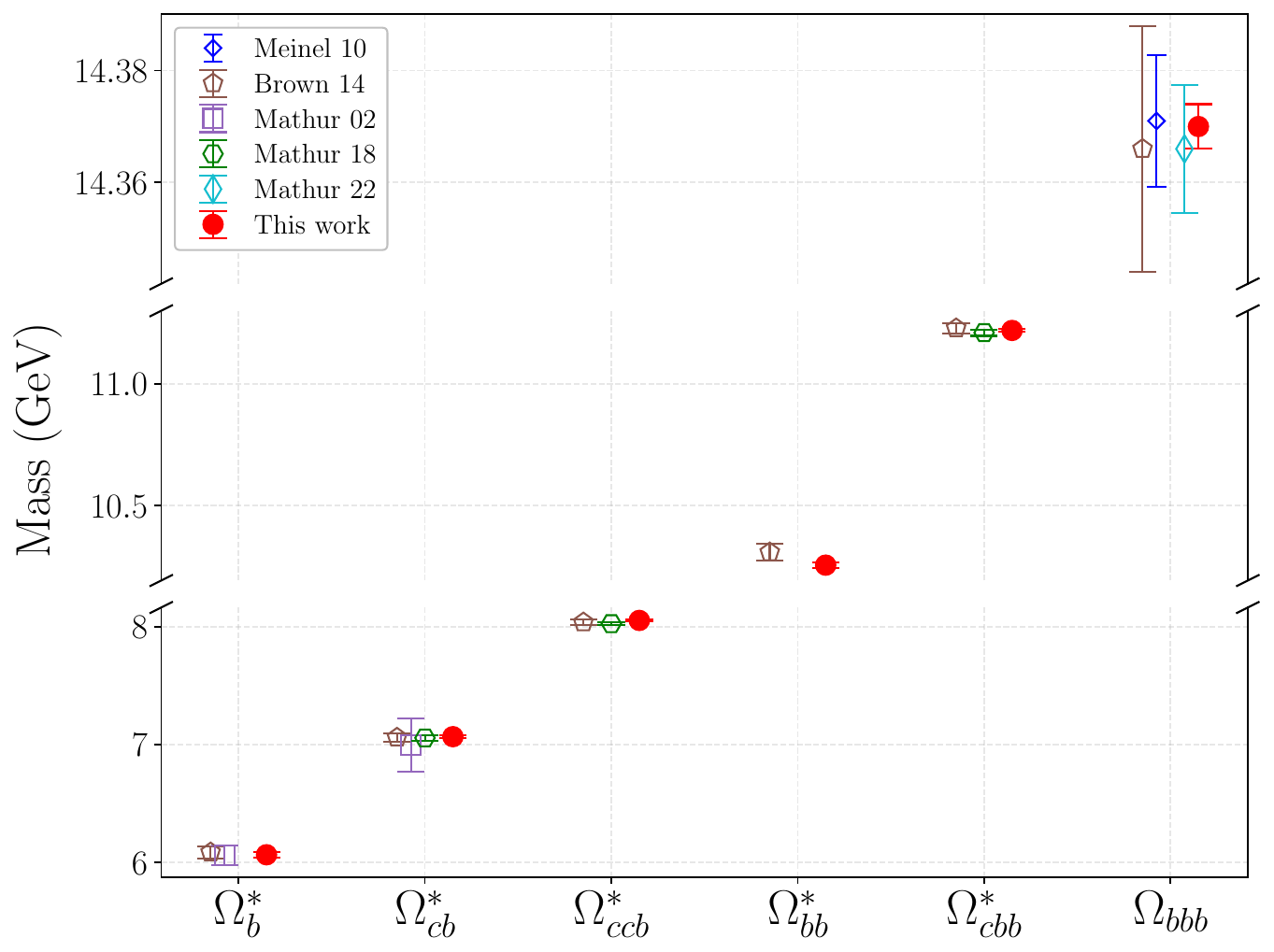}
    \caption{\label{fig:bottom_summary} 
    Summary of the ground-state masses for the bottom and mixed bottom-charm $\Omega$ baryon family ($\Omega_b^*$, $\Omega_{cb}^*$, $\Omega_{ccb}^*$, $\Omega_{bb}^*$, $\Omega_{cbb}^*$, $\Omega_{bbb}$). Our fully relativistic HISQ results (red circles) are compared against previous non-relativistic (NRQCD) calculations by Brown \etal (2014), Mathur \etal (2002,2018 and 2022), showing excellent consistency across the entire heavy quark mass range.}
\end{figure}

We extend this relativistic methodology to the much heavier bottom and mixed bottom-charm sectors, summarized in Fig.~\ref{fig:bottom_summary}. This includes predictions for the ground-state masses of the $\Omega_b^*$, $\Omega_{cb}^*$, $\Omega_{ccb}^*$, $\Omega_{bb}^*$, $\Omega_{cbb}^*$, and the triply heavy $\Omega_{bbb}$. 

A direct comparison reveals that our fully relativistic HISQ predictions are remarkably consistent with Non-Relativistic QCD (NRQCD) calculations by Brown \etal~\cite{Brown:2014ena}, and Mathur \etal~\cite{PhysRevD.66.014502, Mathur:2018epb, PhysRevLett.130.111901}. Because our approach avoids the systematic truncation errors, inherent to the $v^2$ expansion of effective field theories like NRQCD, the observed agreement provides a robust, non-trivial validation of the NRQCD framework while demonstrating that our fully relativistic action has true potential to cleanly resolve exciting physics associated with bottom quarks.


\subsection{Taste splitting and discretization effects}
To explicitly assess residual taste-symmetry breaking in our heavy-baryon spectrum, we computed the ground-state effective masses using independent staggered interpolating operators corresponding to distinct taste constructions. These operators are built using gauge-covariant point-split fields, where different choices of spatial link displacements project onto different taste representations. Specifically, we evaluated the two configurations:
\begin{equation}
    \mathcal{O}_1(t) = \sum_{\mathbf{x} \in \text{Even}} \frac{1}{36} \sum_{\tau, \sigma \in S_3} \epsilon_{abc} \, 
    Q_a^{f_{\sigma(1)}}(\mathbf{x} + \mathbf{a}_{\tau(1)}, t) \,
    D_{\tau(2)} Q_b^{f_{\sigma(2)}}(\mathbf{x} + \mathbf{a}_{\tau(1)}, t) \,
    D_{\tau(3)} Q_c^{f_{\sigma(3)}}(\mathbf{x} + \mathbf{a}_{\tau(1)}, t),
\end{equation}
and
\begin{equation}
    \mathcal{O}_2(t) = \sum_{\mathbf{x} \in \text{Even}} \frac{1}{36} \sum_{\tau, \sigma \in S_3} \epsilon_{abc} \, 
    D_{\tau(1)} Q_a^{f_{\sigma(1)}}(\mathbf{x}, t) \,
    D_{\tau(2)} Q_b^{f_{\sigma(2)}}(\mathbf{x}, t) \,
    D_{\tau(3)} Q_c^{f_{\sigma(3)}}(\mathbf{x}, t),
\end{equation}
where $D_{\tau(i)}$ denotes a product of gauge links implementing spatial displacements in the $\tau(i)$ direction, $S_3$ is the symmetric group denoting permutations over the flavors and spatial directions, and the base sum is restricted to even spatial sites. 
\begin{figure}[htbp]
    \centering
        \includegraphics[width=0.48\textwidth]{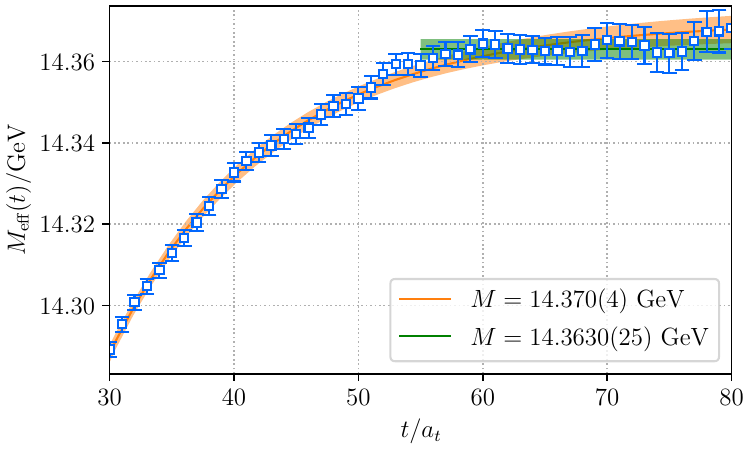}
        \includegraphics[width=0.48\textwidth]{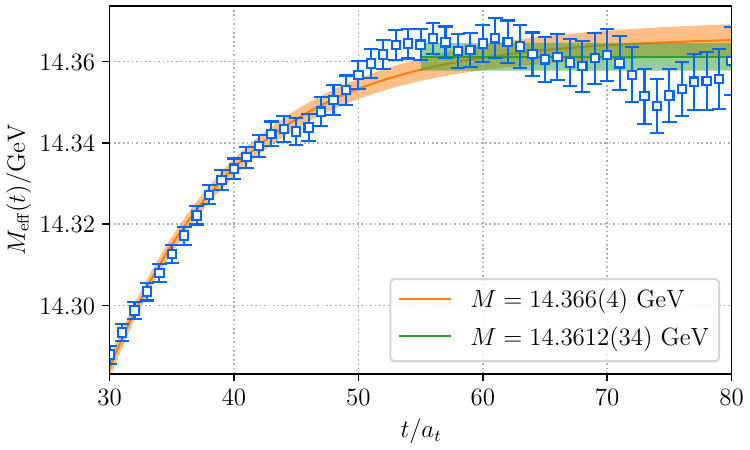}
    \caption{\label{fig:taste_splitting} 
    Effective-mass plateaus, $M_{\text{eff}}(t)$, for the ground state of spin-3/2 $\Omega_{bbb}$ baryon evaluated using two independent staggered interpolating operators with distinct point-split taste constructions. The orange curve indicates the ground-state masses extracted via multi-exponential fits, with shaded bands representing $1\sigma$ statistical uncertainties. The green band denotes the selected fit window for constant fit. The extracted masses, $M = 14.3612(34)$~GeV (right) and $M = 14.3630(25)$~GeV (left), are completely consistent within errors, explicitly demonstrating the expected suppression of taste-breaking effects in the HISQ formalism.}
\end{figure}

As illustrated in Fig.~\ref{fig:taste_splitting}, the correlators constructed from two distinct taste configurations yield statistically indistinguishable ground-state effective mass plateaus. In the HISQ formalism, tree-level $\mathcal{O}(a^2)$ taste-exchange interactions are eliminated, and residual one-loop taste-breaking artifacts of $\mathcal{O}(\alpha_s a^2)$ are heavily suppressed~\cite{Follana:2006rc}. The consistency of our extracted masses — yielding a mass difference of less than $2$~MeV, well within our statistical uncertainties of $\mathcal{O}(3\text{--}4)$~MeV — demonstrates that taste splittings for these heavy baryons are consistent with zero.
\section{\label{sec:comp}Comparison with NRQCD}
Our fully relativistic HISQ approach exhibits improved systematic control over discretization effects compared to traditional methodologies. All previous lattice calculations on baryons involving one or more bottom quarks relied on Non-Relativistic QCD (NRQCD)~\cite{Brown:2014ena,Meinel:2010pw,PhysRevLett.130.111901,PhysRevD.66.014502}. While effective, NRQCD is fundamentally limited by the inability to take  $a \to 0$ continuum limit due to power-law divergences. In contrast, utilization of the HISQ action for bottom quarks  avoids such an issue entirely and permits a systematic continuum extrapolation as in any other traditional lattice calculations. 
\begin{figure}[htbp]
    \centering
        \includegraphics[width=0.7\textwidth]{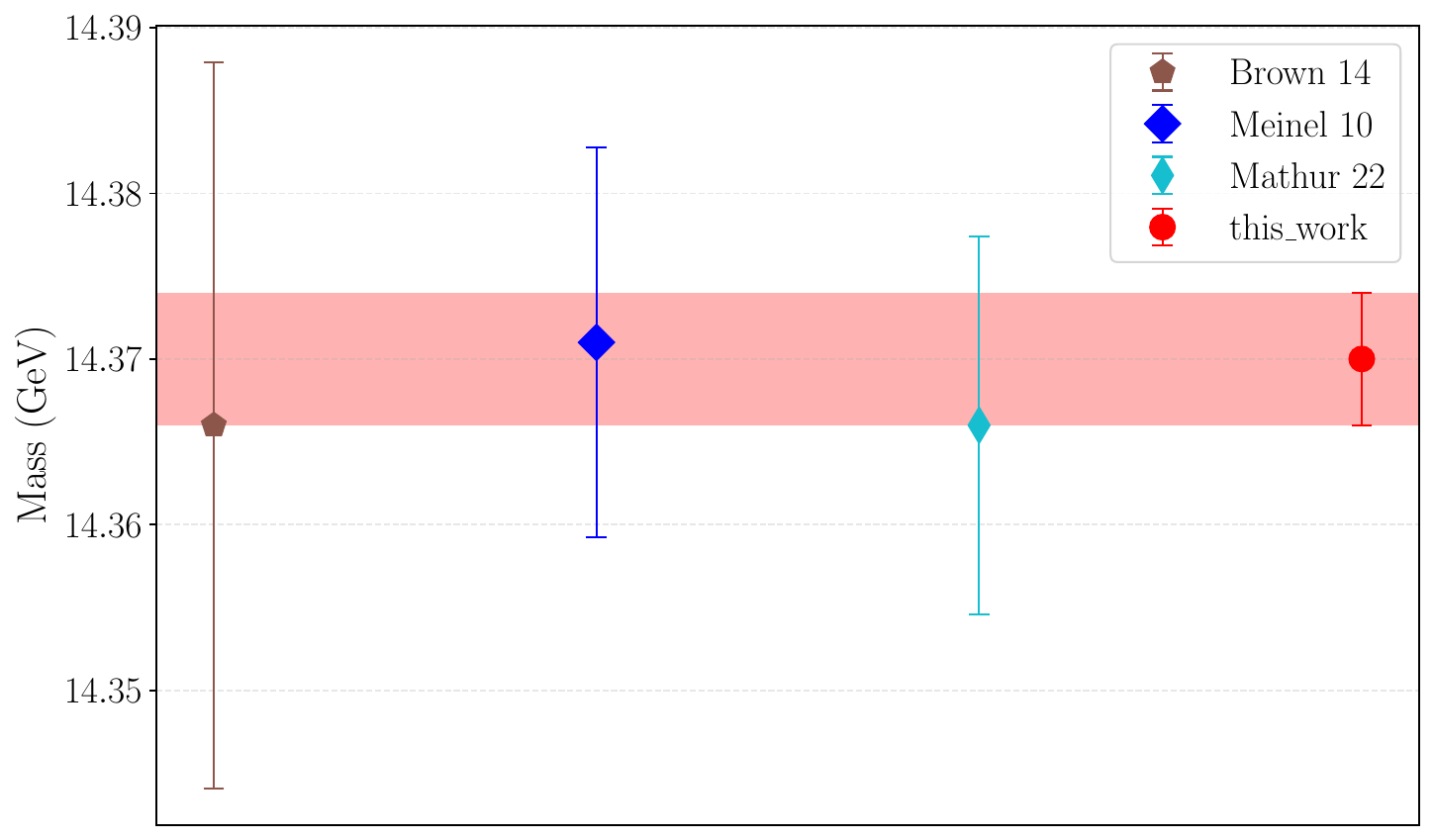}
    \caption{\label{fig:nrqcd_comparison}
    Comparison of the $\Omega_{bbb}(3/2^+)$ ground-state mass extracted from this work with previous lattice determinations. The red marker and the horizontal band represents our fully relativistic measurement using the HISQ action, with the width indicating the $1\sigma$ statistical uncertainty. Systematic uncertainties, including electromagnetic corrections, are not currently included in this calculation, although we expect them to be small. The data points show previous determinations by Brown (2014)~\cite{Brown:2014ena}, Meinel (2010)~\cite{Meinel:2010pw}, and Mathur (2022)~\cite{PhysRevLett.130.111901}, all of which utilize Non-Relativistic QCD for the bottom quark.}
\end{figure}
Despite these profound methodological differences, a direct comparison between relativistic HISQ and NRQCD results demonstrates
consistency for ground-state energies. This can be seen 
in Fig. \ref{fig:bottom_summary} for singly, doubly and triply bottom baryons. In  Fig.~\ref{fig:nrqcd_comparison} we emphasize this further. This agreement provides a nontrivial validation
of NRQCD for heavy baryons while highlighting the advantages of a fully relativistic action
for precise studies in future.

\section{\label{sec:out}Conclusions}

We have presented the {\it first} Lattice QCD study of heavy baryons using a {\it fully} relativistic treatment of all quark flavors. 
We find the taste splittings for heavy baryons are quite negligible. Interestingly, our results also show good agreement with NRQCD-based calculations for these baryons bringing consistency of lattice results using two very different approaches. Future work will focus on improving statistics, performing a detailed systematic error analysis, and extending the study to spin-$1/2$ baryons and excited states. These results open the door to precision studies of heavy-baryon physics using relativistic actions.


\acknowledgments
This work is supported by the Department of Atomic Energy, Government of India, under Project Identification Number RTI-4012.
Computations were carried out on the computing clusters at the Department of Theoretical Physics, TIFR, Mumbai. We are thankful to the MILC collaboration and, in particular, to S. Gottlieb for providing us with the HISQ lattice ensembles. We would also like to thank  Ajay Salve and Kapil Ghadiali for computational support.


\bibliographystyle{JHEP}
\bibliography{proc}

@article{Follana:2006rc,
    author = "Follana, E. and others",
    title = "{Highly Improved Staggered Quarks on the Lattice, with Applications to Charm Physics}",
    eprint = "hep-lat/0610092",
    archivePrefix = "arXiv",
    doi = "10.1103/PhysRevD.75.054502",
    journal = "Phys. Rev. D",
    volume = "75",
    pages = "054502",
    year = "2007"
}

@article{Dhindsa:2024erk,
    author = "Dhindsa, Navdeep Singh and Chakraborty, Debsubhra and Radhakrishnan, Archana and Mathur, Nilmani and Padmanath, M.",
    title = "{Precise study of triply charmed baryons {\ensuremath{\Omega}}ccc}",
    eprint = "2411.12729",
    archivePrefix = "arXiv",
    primaryClass = "hep-lat",
    reportNumber = "TIFR/TH/24-22",
    doi = "10.1103/9phq-k8pv",
    journal = "Phys. Rev. D",
    volume = "112",
    number = "11",
    pages = "L111501",
    year = "2025"
}

@article{Hatton:2020qhk,
    author = "Hatton, D. and Davies, C. T. H. and Galloway, B. and Koponen, J. and Lepage, G. P. and Lytle, A. T.",
    collaboration = "HPQCD",
    title = "{Charmonium properties from lattice $QCD$+QED : Hyperfine splitting, $J/\psi$ leptonic width, charm quark mass, and $a^c_\mu$}",
    eprint = "2005.01845",
    archivePrefix = "arXiv",
    primaryClass = "hep-lat",
    doi = "10.1103/PhysRevD.102.054511",
    journal = "Phys. Rev. D",
    volume = "102",
    number = "5",
    pages = "054511",
    year = "2020"
}

@article{Bazavov:2012xda,
    author = "Bazavov, A. and others",
    collaboration = "MILC",
    title = "{Lattice QCD ensembles with four flavors of highly improved staggered quarks}",
    eprint = "1212.4768",
    archivePrefix = "arXiv",
    primaryClass = "hep-lat",
    doi = "10.1103/PhysRevD.87.054505",
    journal = "Phys. Rev. D",
    volume = "87",
    pages = "054505",
    year = "2013"
}

@article{Mattson:2002vu,
    author = "Mattson, M. and others",
    collaboration = "SELEX",
    title = "{First Observation of the Doubly Charmed Baryon $\Xi_{cc}^+$}",
    eprint = "hep-ex/0208014",
    archivePrefix = "arXiv",
    doi = "10.1103/PhysRevLett.89.112001",
    journal = "Phys. Rev. Lett.",
    volume = "89",
    pages = "112001",
    year = "2002"
}

@article{Aaij:2017ueg,
    author = "Aaij, R. and others",
    collaboration = "LHCb",
    title = "{Observation of the doubly charmed baryon $\Xi_{cc}^{++}$}",
    eprint = "1707.01621",
    archivePrefix = "arXiv",
    primaryClass = "hep-ex",
    doi = "10.1103/PhysRevLett.119.112001",
    journal = "Phys. Rev. Lett.",
    volume = "119",
    number = "11",
    pages = "112001",
    year = "2017"
}

@article{Brown:2014ena,
    author = "Brown, Zachary S. and Detmold, William and Meinel, Stefan and Orginos, Kostas",
    title = "{Charmed bottom baryon spectroscopy from lattice QCD}",
    eprint = "1409.0497",
    archivePrefix = "arXiv",
    primaryClass = "hep-lat",
    doi = "10.1103/PhysRevD.90.094507",
    journal = "Phys. Rev. D",
    volume = "90",
    number = "9",
    pages = "094507",
    year = "2014"
}

@article{Meinel:2010pw,
    author = "Meinel, Stefan",
    title = "{Prediction of the $\Omega_{bbb}$ mass from lattice QCD}",
    eprint = "1008.3154",
    archivePrefix = "arXiv",
    primaryClass = "hep-lat",
    doi = "10.1103/PhysRevD.82.114514",
    journal = "Phys. Rev. D",
    volume = "82",
    pages = "114514",
    year = "2010"
}

@article{Briceno:2012wt,
    author = "Briceno, Raul A. and Lin, Huey-Wen and Bolton, Daniel R.",
    title = "{Charmed-Baryon Spectroscopy from Lattice QCD with $N_f$ = 2+1+1 Flavors}",
    eprint = "1207.3536",
    archivePrefix = "arXiv",
    primaryClass = "hep-lat",
    reportNumber = "NT-UW-12-12, NT@UW-12-12",
    doi = "10.1103/PhysRevD.86.094504",
    journal = "Phys. Rev. D",
    volume = "86",
    pages = "094504",
    year = "2012"
}

@article{Mathur:2018epb,
    author = "Mathur, Nilmani and Padmanath, M. and Mondal, Sourav",
    title = "{Precise predictions of charmed-bottom hadrons from lattice QCD}",
    eprint = "1806.04151",
    archivePrefix = "arXiv",
    primaryClass = "hep-lat",
    reportNumber = "TIFR/TH/18-09, TIFR-TH-18-09",
    doi = "10.1103/PhysRevLett.121.202002",
    journal = "Phys. Rev. Lett.",
    volume = "121",
    number = "20",
    pages = "202002",
    year = "2018"
}

@article{Mathur:2018rwu,
    author = "Mathur, Nilmani and Padmanath, M.",
    title = "{Lattice QCD study of doubly-charmed strange baryons}",
    eprint = "1807.00174",
    archivePrefix = "arXiv",
    primaryClass = "hep-lat",
    reportNumber = "TIFR/TH/18-18, TIFR-TH-18-18",
    doi = "10.1103/PhysRevD.99.031501",
    journal = "Phys. Rev. D",
    volume = "99",
    number = "3",
    pages = "031501",
    year = "2019"
}

@article{Alexandrou:2023dlu,
    author = "Alexandrou, Constantia and Bacchio, Simone and Christou, Georgios and Finkenrath, Jacob",
    title = "{Low-lying baryon masses using twisted mass fermions ensembles at the physical pion mass}",
    eprint = "2309.04401",
    archivePrefix = "arXiv",
    primaryClass = "hep-lat",
    doi = "10.1103/PhysRevD.108.094510",
    journal = "Phys. Rev. D",
    volume = "108",
    number = "9",
    pages = "094510",
    year = "2023"
}

@article{RQCD:2022xux,
    author = {Bali, Gunnar S. and Collins, Sara and Georg, Peter and Jenkins, Daniel and Korcyl, Piotr and Sch{\"a}fer, Andreas and Scholz, Enno E. and Simeth, Jakob and S{\"o}ldner, Wolfgang and Weish{\"a}upl, Simon},
    collaboration = "RQCD",
    title = "{Scale setting and the light baryon spectrum in N$_{f}$ = 2 + 1 QCD with Wilson fermions}",
    eprint = "2211.03744",
    archivePrefix = "arXiv",
    primaryClass = "hep-lat",
    doi = "10.1007/JHEP05(2023)035",
    journal = "JHEP",
    volume = "05",
    pages = "035",
    year = "2023"
}

@article{Hu:2024mas,
    author = "Hu, Bolun and Du, Haiyang and Jiang, Xiangyu and Liu, Keh-Fei and Sun, Peng and Yang, Yi-Bo",
    title = "{Unveiling the Strong Interaction origin of Baryon Masses with Lattice QCD}",
    eprint = "2411.18402",
    archivePrefix = "arXiv",
    primaryClass = "hep-lat",
    month = "11",
    year = "2024"
}

@article{BMW:2008jgk,
    author = "Durr, S. and others",
    collaboration = "BMW",
    title = "{Ab-Initio Determination of Light Hadron Masses}",
    eprint = "0906.3599",
    archivePrefix = "arXiv",
    primaryClass = "hep-lat",
    doi = "10.1126/science.1163233",
    journal = "Science",
    volume = "322",
    pages = "1224--1227",
    year = "2008"
}

@article{PhysRevD.64.094509,
  title = {Charmed baryons in lattice QCD},
  author = {Lewis, Randy and Mathur, N. and Woloshyn, R. M.},
  journal = {Phys. Rev. D},
  volume = {64},
  issue = {9},
  pages = {094509},
  numpages = {9},
  year = {2001},
  month = {Oct},
  publisher = {American Physical Society},
  doi = {10.1103/PhysRevD.64.094509},
  url = {https://link.aps.org/doi/10.1103/PhysRevD.64.094509}
}

@article{PhysRevLett.130.111901,
  title = {Strongly Bound Dibaryon with Maximal Beauty Flavor from Lattice QCD},
  author = {Mathur, Nilmani and Padmanath, M. and Chakraborty, Debsubhra},
  journal = {Phys. Rev. Lett.},
  volume = {130},
  issue = {11},
  pages = {111901},
  numpages = {7},
  year = {2023},
  month = {Mar},
  publisher = {American Physical Society},
  doi = {10.1103/PhysRevLett.130.111901},
  url = {https://link.aps.org/doi/10.1103/PhysRevLett.130.111901}
}

@article{PhysRevD.66.014502,
  title = {Charmed and bottom baryons from lattice nonrelativistic QCD},
  author = {Mathur, Nilmani and Lewis, Randy and Woloshyn, R. M.},
  journal = {Phys. Rev. D},
  volume = {66},
  issue = {1},
  pages = {014502},
  numpages = {10},
  year = {2002},
  month = {Jul},
  publisher = {American Physical Society},
  doi = {10.1103/PhysRevD.66.014502},
  url = {https://link.aps.org/doi/10.1103/PhysRevD.66.014502}
}

@article{PhysRevD.86.014501,
  title = {${\ensuremath{\Delta}}_{\mathrm{mix}}$ parameter in the overlap on domain-wall mixed action},
  author = {Lujan, M. and Alexandru, A. and Chen, Y. and Draper, T. and Freeman, W. and Gong, M. and Lee, F. X. and Li, A. and Liu, K. F. and Mathur, N.},
  collaboration = {QCD Collaboration},
  journal = {Phys. Rev. D},
  volume = {86},
  issue = {1},
  pages = {014501},
  numpages = {6},
  year = {2012},
  month = {Jul},
  publisher = {American Physical Society},
  doi = {10.1103/PhysRevD.86.014501},
  url = {https://link.aps.org/doi/10.1103/PhysRevD.86.014501}
}

@article{Basak:2012py,
    author = "Basak, S. and Datta, S. and Padmanath, M. and Majumdar, P. and Mathur, N.",
    editor = "Leinweber, Derek and Kamleh, Waseem and Mahbub, Selim and Matevosyan, Hrayr and Thomas, Anthony and Williams, Anthony G. and Young, Ross and Zanotti, James",
    title = "{Charm and strange hadron spectra from overlap fermions on HISQ gauge configurations}",
    eprint = "1211.6277",
    archivePrefix = "arXiv",
    primaryClass = "hep-lat",
    reportNumber = "TIFR-TH-12-44",
    doi = "10.22323/1.164.0141",
    journal = "PoS",
    volume = "LATTICE2012",
    pages = "141",
    year = "2012"
}

@article{DeTar:2014gla,
    author = "DeTar, Carleton and Lee, Song-Haeng",
    title = "{Variational method with staggered fermions}",
    eprint = "1411.4676",
    archivePrefix = "arXiv",
    primaryClass = "hep-lat",
    doi = "10.1103/PhysRevD.91.034504",
    journal = "Phys. Rev. D",
    volume = "91",
    number = "3",
    pages = "034504",
    year = "2015"
}
\end{document}